\documentclass[aps,a4paper, preprint, superscriptaddress,preprintnumbers,floatfix,nofootinbib,amsmath,amssymb]{revtex4}
\usepackage{url}
\usepackage{hyperref}
\usepackage{cleveref}

\usepackage{amstext,amssymb}
\usepackage{amsmath,bm}
\usepackage{graphicx}
\usepackage{xspace}
\usepackage{color}
\usepackage{units}

\usepackage{feynmp}
\usepackage{cancel}
\usepackage{bm}
\usepackage{slashed} 
\usepackage{multirow}
\newcommand{\be}{\begin{equation}}
\newcommand{\ee}{\end{equation}}
\newcommand{\bea}{\begin{eqnarray}}
\newcommand{\eea}{\end{eqnarray}}

\def\s1{\hat s}

\usepackage{slashed}

\newcommand{\nua}[1]{\ensuremath{\rlap{\kern-2.5pt\ensuremath{\overset{\scriptscriptstyle(-)}{\phantom{\nu}}}}{\ensuremath{{\nu}_{#1}}}}\xspace}

\begin{document}
\title{ Phenomenology of inverse seesaw using  $S_3$ modular symmetry}

\author{Mitesh Kumar Behera}
\email{miteshbehera1304@gmail.com}
\affiliation{High Energy Physics Research Unit, Department of Physics, Faculty of Science, Chulalongkorn University, Bangkok 10330, Thailand}

\author{Pawin Ittisamai}
\email{pawin.i@chula.ac.th}
\affiliation{High Energy Physics Research Unit, Department of Physics, Faculty of Science, Chulalongkorn University, Bangkok 10330, Thailand}

\author{Chakrit Pongkitivanichkul}
\email{chakpo@kku.ac.th}
\affiliation{Khon Kaen Particle Physics and Cosmology Theory Group (KKPaCT), Department of Physics, Faculty of Science, Khon Kaen University, 123 Mitraphap Rd, Khon Kaen 40002, Thailand}

\author{Patipan Uttayarat}
\email{patipan@g.swu.ac.th}
\affiliation{Theoretical High-Energy Physics and Astrophysics Research Unit (ThEPA), Department of Physics, Srinakharinwirot University, 114 Sukhumvit 23 Rd., Wattana, Bangkok
10110, Thailand}

\begin{abstract}
Describing neutrino masses using the inverse seesaw mechanism with discrete flavor symmetry imposed through modular forms provides a testable framework at TeV scales with fewer parameters. However, $S_3$, the smallest modular group, remains relatively underexplored. In this work, we construct the minimal supersymmetric inverse seesaw model based on the modular $S_3$ flavor symmetry. In our model, the light neutrino mass matrix depends on 6 real parameters: the complex modulus, an overall scale for light neutrino mass, a real ratio and a complex ratio of Yukawa coupling. Thanks to its minimality, our model offers various definite predictions: the lightest neutrino is massless, the neutrino masses are inverted ordering, the sum of the three light neutrino masses ($\sum_i m_i$) is 100 meV, the effective mass for the end point of the beta decay spectrum is 50 meV, the effective mass for neutrinoless double beta decay ($m_{ee}$) is in the range $38-58$ meV. In particular, the predicted values for $\sum_i m_i$ and $m_{ee}$ from our model are within reach of the next generation experiments. Our model also predicts radiative lepton flavor violating decays $\ell\to\ell'\gamma$ which are compatible with experimental constraints.

\end{abstract}

\maketitle
\flushbottom

\section{Introduction}
The Standard Model (SM) originally described neutrinos as massless particles. However, experimental evidence from neutrino oscillation data has confirmed that neutrinos have small but non-zero masses. This discovery highlights the phenomenon of neutrino mixing and implies that at least two neutrino mass states are non-zero~\cite{King:2003jb, Altarelli:2004za, Mohapatra:2006gs}. Unlike other fermions in the SM, neutrinos lack right-handed components, preventing them from obtaining masses via the Higgs mechanism. Instead, the dimension-five Weinberg operator~\cite{Weinberg:1979sa, Wilczek:1979hc, Weinberg:1980bf} offers a plausible explanation for neutrino mass generation, though its origin and flavor structure remain unresolved. Consequently, extending the SM with Beyond Standard Model (BSM) frameworks is essential to account for these masses.

One of the simplest and most studied UV completion of the Weinberg operator is the so called type-I seesaw model where the SM is extended by a set of electroweak singlet right-handed neutrinos~\cite{Minkowski:1977sc, Mohapatra:1979ia, Gell-Mann:1979vob, Yanagida:1979as, Glashow:1979nm, Chen:2009um}. In this framework, the mass of the light neutrinos are $m_\nu\sim v^2/M_R$ where $v=246$ GeV is the electroweak vacuum expectation value (VEV) and $M_R$ is the Majorana mass of the right-handed neutrino. Small neutrino mass of order 0.1 eV requires $M_R\sim10^{13}$ GeV, rendering it impossible to probe such a scenario in the current and planned collider experiments. 
Several models have been proposed to lower the new physics scale responsible for neutrino mass generation. They include the type-II seesaw models~\cite{Magg:1980ut,Schechter:1980gr,Cheng:1980qt,Lazarides:1980nt,Mohapatra:1980yp,Primulando:2019evb}, the radiative neutrino mass models~\cite{Zee:1980ai, Zee:1985id, Babu:1988ki, Babu:2013pma, Nomura:2019dhw, Primulando:2022lpj}, and models involving extra dimensions~\cite{Arkani-Hamed:1998wuz, Carena:2017qhd, Forero:2022skg, Roy:2023dyq, Anchordoqui:2023wkm}. 
Another interesting variant of the seesaw models is the inverse seesaw mechanism~\cite{Mohapatra:1986bd,Malinsky:2005bi,Dias:2012xp,Dias:2011sq,Abada:2014vea,Nomura:2018cfu,Pongkitivanichkul:2019cvm,Thongyoi:2025ghi,Panda:2022kbn}. In this framework, the SM is extended by a set of electroweak singlet left-handed and right-handed neutrinos. Both the left- and right-handed sterile neutrinos admit small, yet technically natural, lepton number violating (LNV) Majorana mass terms. The presence of these small LNV parameters plays a crucial role in bringing the seesaw scale down to the TeV scale. 

The generic models of neutrino mass usually suffer from the large number of free parameters in the Yukawa sector, make it impossible to get definite predictions out of them. To address this problem, various discrete symmetries have been introduced in the literature~\cite{Altarelli:2005yp, Altarelli:2010gt, Ishimori:2010au, Hernandez:2012ra, King:2013eh, King:2014nza, Hagedorn:2017zks, Patel:2023voj}. The discrete flavor symmetry is imposed with the help of flavon fields. As a result, such models bring with them extra challenges of aligning the VEV of the flavons to ensure acceptable neutrino masses and mixings.

Recently, it has been proposed that discrete flavor symmetry can be imposed with the help of modular forms~\cite{Leontaris:1997vw, Kobayashi:2018vbk, Feruglio:2019neutrino, deAdelhartToorop:2011re}. In this framework, Yukawa couplings and Majorana mass parameters are elevated to modular forms, which transform under specific representations of the discrete flavor symmetry group. This approach has been widely applied to construct viable neutrino mass models with symmetry groups like \(S_3\)~\cite{Kobayashi:2018wkl, Okada:2019xqk, Mishra:2020gxg,Meloni:2023aru,Behera:2024ark,Marciano:2024nwm, Belfkir:2024uvj,Okada:2025jjo}, \(A_4\)~\cite{Nomura:2023usj, RickyDevi:2024ijc, Kumar:2023moh, Mishra:2022egy,Gogoi:2023jzl,Pathak:2024sei, Wang:2019xbo,Nomura:2024vus,Kashav:2021zir, Kashav:2022kpk, Mishra:2023ekx, Lu:2019vgm, Kobayashi:2019gtp, Nomura:2019xsb, Behera:2020lpd, MiteshKumar:2023hpg, Nomura:2023kwz, Kim:2023jto, Devi:2023vpe, Dasgupta:2021ggp, Nomura:2019lnr, CentellesChulia:2023osj}, \(S_4\)~\cite{Penedo:2018nmg, Novichkov:2018ovf, Kobayashi:2019xvz, Liu:2020akv, deMedeirosVarzielas:2023crv, Ding:2021zbg, King:2019vhv}, and \(A_5\)~\cite{Novichkov:2018nkm, Ding:2019zxk}, as well as more extensive groups~\cite{Baur2019Unification}. Further studies have explored the double coverings of \(A_4\)~\cite{Mishra:2023cjc, Liu:2019khw, Ding:2023ynd}, \(S_4\)~\cite{Abe:2023ilq, Abe:2023qmr}, and \(A_5\)~\cite{Wang:2020lxk, Behera:2022wco, Behera:2021eut}. Among these, \(S_3\), the smallest finite modular group, remains relatively underexplored.

In this work, we investigate the inverse seesaw model in the context of the modular $S_3$ flavor symmetry. To naturally impose modular symmetry, we make our model supersymmetric. We identify the minimal Yukawa sector which is compatible with neutrino oscillation data. The minimality of our model allows us to make definite predictions regarding neutrino masses and related effective masses. We also study constraints imposed by lepton flavor violation (LFV) on the models. 

The manuscript is organized as follows: in Section~\ref{sec:model}, we introduce the model and derive the neutrino mass matrices. Neutrino phenomenology is discussed in Section~\ref{sec:neut pheno}. In Section~\ref{sec:lfv} we work out lepton flavor violation observables induced in our model. In Section~\ref{sec:results} we present our numerical analysis consistent with experimental data. Finally, we offer our conclusion and discussion in Section~\ref{sec:con}.

\section{The model}
\label{sec:model}
The leptonic sector of the model consists of the electroweak doublet chiral superfield $L_i$, the charged lepton chiral superfield $E^c_i$ ($i=1,2,3$) and the electroweak singlet chiral superfield $N^c_I$ and $S_I$ ($I=1,2$). The superfield $(L_1,L_2)$, $(E^c_1,E^c_2)$, $(N^c_1,N^c_2)$ and $(S_1,S_2)$ are grouped into doublets of $S_3$, while the superfields $L_3$ and $E^c_3$ are singlets under $S_3$. For notational compactness, we introduce the shorthand notation $X\equiv(X_1,X_2)$ for the $S_3$ doublet. The Higgs superfields $H_u$ and $H_d$ are electroweak doublets with hypercharge $+1/2$ and $-1/2$ respectively. The transformation properties under $SU(2)_L\times U(1)_Y\times S_3$ and the corresponding modular weight of the chiral superfields are summarized in Tab.~\ref{tab:fieldcontent}.

\begin{center} 
\begin{table}[h!]
\centering
\begin{tabular}{|c||c c|c c|c c|c|}\hline\hline  
& $E^c$& $E_{3}^c$ & $L$& $L_3$& $N^c$&  $S$& $H_{u,d}$  \\ \hline \hline
$SU(2)_L$ & 1  & $1$  & $2$  & $2$  & $1$  &   $1$   & $2$        \\\hline 
$U(1)_Y$  & $1$  & $1$   & $-\frac{1}{2}$      & $-\frac{1}{2}$      & $0$       & $0$      & $\pm \frac{1}{2}$    \\\hline 
$S_3$     & $2$  & $1$  & $2$  & $1$  & $2$  & $2$  & $1$    \\ \hline
$k_I$     & $-1$       & $-3$       & $3$       & $3$       & $1$      &  $1$   & $0$               \\ \hline
\end{tabular}
\caption{Transformation properties of the superfields under $SU(2)_L\times U(1)_Y\times S_3$ and their modular weight ($k_I$).}
\label{tab:fieldcontent}
\end{table}
\end{center}

The super potential in the leptonic sector consistent with the modular $S_3$ symmetry is given by
\begin{equation}
\begin{aligned}
    \mathcal{W}_\ell &= \alpha_\ell (E^c L)_2 Y_2^{(2)} H_d  -i\beta_\ell (E^c Y_2^{(2)}) L_3 H_d + \gamma_\ell E_3^c L_3 H_d\\
    &\quad+\alpha_D (N^c L)_2 Y_2^{(4)} H_u + \gamma_D (N^c L) Y_1^{(4)} H_u - i\beta_D (N^c Y_2^{(4)}) L_3 H_u \\ 
    &\quad + M (N^cS)_2  Y_2^{(2)}  +\frac{\mu_N}{2} (N^c N^c)_2 Y_2^{(2)} + \frac{\mu_S}{2} (S S)_2 Y_2^{(2)},
\end{aligned}
\label{eq:superpotential}
\end{equation}
where $(\ldots)_a$ denotes the $S_3$ charge, $Y_a^{(k)}$ denotes the modular form with the $S_3$ charge $a$ and modular weight $k$. The $S_3$ product rule and the modular form $Y_a^{(k)}$ up to modular weight 4 are listed in App.~\ref{sec:S3_symmetry}. Note that the factors $-i$ in front of $\beta_\ell$ and $\beta_D$ are for later convenience. The parameters $\alpha$'s, $\beta$'s and $\gamma$'s are complex in general. However, we can absorb the phases of $\alpha_L$, $\beta_L$, $\gamma_L$, $\alpha_D$ and $\beta_D$ into $L$, $E^c_{\phantom{i}}$, $E^c_3$, $N^c$ and $L_3$ respectively, leaving $\gamma_D$ the only complex parameter beside the modular form. 

The first line in Eq.~\eqref{eq:superpotential}, after electroweak symmetry breaking, leads to the charged lepton mass matrix $(M_\ell)_{ij}E^c_iL^{\phantom{c}}_j$, while the second line gives rise to the Dirac mass matrix $(M_D)_{Ij}N^c_IL^{\phantom{c}}_j$. The last line leads to Majorana mass matrices for $N^c_I$ and $S_I^{\phantom{c}}$, $\frac{1}{2}(M_R)_{IJ}N_I^cN_J^c$ + $\frac{1}{2}(M_S)_{IJ}S_IS_J$, as well as the Dirac mass matrix $(M_{NS})_{IJ}N_I^cS_J^{\phantom{c}}$. Before we analyze the particle spectrum of the model, we note that by performing unitary transformation on the $S_3$ doublets superfields
\begin{equation}
    \psi \to \frac{-i}{\sqrt{2}}\begin{pmatrix}1 &-1\\i &i\end{pmatrix}\psi,
\end{equation}
where $\psi=L,E^c,N^c,S$. The mass matrices can be put in the form
\begin{align}
    M_\ell &= \frac{v_d}{\sqrt{2}}\begin{pmatrix}
        \alpha_\ell Y_-  & 0 & \frac{-\beta_\ell}{\sqrt{2}} Y_+\\
        0 & \alpha_\ell Y_+ & \frac{\beta_\ell}{\sqrt{2}} Y_- \\
        0  & 0 & \gamma_\ell\end{pmatrix},\\
    M_D &= \frac{v_u}{\sqrt{2}}\begin{pmatrix}
        -\alpha_D Y_+^2 &\gamma_D Y_-Y_+ & \frac{\beta_D}{\sqrt{2}} Y_-^2\\
        \gamma_D Y_-Y_+ & -\alpha_D Y_-^2 & -\frac{\beta_D}{\sqrt{2}} Y_+^2\end{pmatrix},\\
    M_R,M_S,M_{NS} &\sim \begin{pmatrix}Y_-&0\\0&Y_+\end{pmatrix},
\end{align}
where $v_{u}$ and $v_d$ are the vacuum expectation values of $H_u$ and $H_d$ respectively with $v=\sqrt{v_u^2+v_d^2} = 246$ GeV. In the above equation is we define $Y_\pm\equiv Y_1\pm iY_2$.

The charged lepton mass matrix, $M_\ell$, depends on three real parameters $\alpha_\ell$, $\beta_\ell$ and $\gamma_\ell$. These parameters must be chosen so that eigenvalues of $M_\ell$ reproduce the observed charged lepton masses. This can be achieved through the help of the invariants
\begin{align}
    \text{Tr}(M_\ell^\dagger M_\ell^{}) &= m_e^2 + m_\mu^2 + m_\tau^2,\label{eq:trace}\\
    \frac{1}{2}\left((\text{Tr}[M_\ell^\dagger M_\ell^{}])^2 - \text{Tr}[(M_\ell^\dagger M_\ell^{})^2]\right) &= m_e^2m_\mu^2 + m_e^2m_\tau^2 + m_\mu^2m_\tau^2,\label{eq:cross}\\
    \det(M_\ell^\dagger M_\ell^{}) &= m_e^2m_\mu^2m_\tau^2\label{eq:det},
\end{align}
where $m_e$, $m_\mu$, and $m_\tau$ are the masses of electron, muon, and tau, respectively. After $\alpha_\ell$, $\beta_\ell$ and $\gamma_\ell$ are determined, the charged lepton mass matrix can be diagonalized by a unitary matrix $U_L$ such that
\begin{equation}
    U_L^\dagger M_\ell^\dagger M_\ell^{}U_L^{} = \text{diag.}(m_e^2,m_\mu^2,m_\tau^2).
    \label{eq:UL}
\end{equation}

In the neutrino sector, we can write the neutrino mass term as
\begin{equation}
    \frac12(M_{\mathcal{N}})_{\mathcal{I}\mathcal{J}}\mathcal{N}_\mathcal{I}\mathcal{N}_\mathcal{J},
\end{equation}
with $\mathcal{N}_\mathcal{I} = (\nu_i^{},N^c_I,S^{}_J)$ and
\begin{equation}
    M_{\mathcal{N}} = \begin{pmatrix}
        0 & M_D^T &0\\
        M_D & M_R &M_{NS}\\
        0 &M_{NS}^T &M_S
    \end{pmatrix}.
\end{equation}
In the inverse seesaw limit where $||M_R||,||M_S||\ll||M_D||\ll||M_{NS}||$, the light neutrino mass matrix is approximately given by
\begin{equation}
\begin{aligned}
    m_\nu &\simeq M_D^TM_{NS}^{-1}M_S^{}M_{NS}^{-1}M_D\\
    &= \kappa\begin{pmatrix}
        \frac{Y_+}{Y_-}(Y_+^3\tilde\alpha_D^2 + Y_-^3\tilde\gamma_D^2)& -(Y_+^3+Y_-^3)\tilde\alpha_D\tilde\gamma_D & -\frac{Y_+^2Y_-}{\sqrt{2}}(\tilde\alpha_D+\tilde\gamma_D)\\
        -(Y_+^3+Y_-^3)\tilde\alpha_D\tilde\gamma_D & \frac{Y_-}{Y_+}(Y_-^3\tilde\alpha_D^2 + Y_+^3\tilde\gamma_D^2)&\frac{Y_+Y_-^2}{\sqrt{2}}(\tilde\alpha_D+\tilde\gamma_D)\\
        -\frac{Y_+^2Y_-}{\sqrt{2}}(\tilde\alpha_D+\tilde\gamma_D)&\frac{Y_+Y_-^2}{\sqrt{2}}(\tilde\alpha_D+\tilde\gamma_D) & \frac12(Y_+^3+Y_-^3)
    \end{pmatrix},
\end{aligned}
\label{eq:numass}
\end{equation}
where
\begin{equation}
    \kappa=\beta_D^2\sin^2\beta\,\frac{\mu_S\,v^2}{M^2}
    \label{eq:kappa}
\end{equation}
is the overall scale of the neutrino mass matrix, $\tan\beta=v_u/v_d$, and $\tilde x = x/\beta_D$. Since $m_\nu$ is rank 2, the lightest neutrino mass in our scenario vanishes. On the other hand, the heavy neutrino mass matrix is 
\begin{equation}
    M_N \simeq \begin{pmatrix}0&M_{NS}\\M_{NS}^T&0\end{pmatrix}.
\end{equation}
Note that the heavy neutrino states are approximately Dirac neutrino with masses $M|Y_\pm|$. 

The light neutrino mass matrix, $m_\nu$, can be diagonalized by a unitary matrix $U_\nu$ with
\begin{equation}
    U_\nu^Tm_\nu U_\nu^{} = \text{diag.}(m_1,m_2,m_3),
    \label{eq:Unu}
\end{equation}
where $m_i$ is real and nonnegative. There are two possible orderings of neutrino masses in Eq.~\eqref{eq:Unu}, the normal ordering (NO) where $m_1<m_2<m_3$ and the inverted ordering (IO) in which $m_3<m_1<m_2$. The matrix $U_\nu$, together with the matrix $U_L$ in Eq.~\eqref{eq:UL}, defines the Pontecorvo-Maki-Nakagawa-Sakata (PMNS)
\begin{equation}
    U = U_L^\dagger U_\nu^{}.
    \label{eq:pmns}
\end{equation}

\section{Neutrino phenomenology}
\label{sec:neut pheno}
Neutrino oscillation data can be described by six parameters: two mass squared differences $\Delta m^2_{\rm sol}$ and $\Delta m^2_{\rm atm}$, three mixing angles $\theta_{12}$, $\theta_{23}$ and $\theta_{13}$, and one CP violating phase $\delta$. After the diagonalization of the charged and light neutrino mass matrices, one can deduce the neutrino oscillation parameters.  For the mass-squared splittings we have $\Delta m^2_{\rm sol} = m_2^2-m_1^2$ for both NO and IO, and $\Delta m^2_{\rm atm} = m_3^2-m_1^2$ and $m_2^2-m_3^2$ for NO and IO respectively. The PMNS matrix, see Eq.~\eqref{eq:pmns}, is characterized by three mixing angles $\theta_{12}$, $\theta_{23}$, $\theta_{13}$, the Dirac CP violating phase $\delta$ and two Majorana phases. After diagonalizing the charged lepton and the light neutrino mass matrices, the sine of the mixing angles are given by
\begin{align}
    \sin^2\theta_{12} &= \frac{|U_{e2}|^2}{1-|U_{e3}|^2},\\
    \sin^2\theta_{23} &= \frac{|U_{\mu3}|^2}{1-|U_{e3}|^2},\\
    \sin^2\theta_{13} &= |U_{e3}|^2.
\end{align}
Meanwhile, the Dirac phase can be deduced from the product $U_{e2}^{}U_{\mu3}^{}U_{e3}^\ast U_{\mu2}^\ast$. Finally, the two Majorana phases have never been experimentally measured. However, in principle, their can be determined from the product $U_{e1}U_{e2}^\ast$ and $U_{e1}U_{e3}^\ast$~\cite{Jenkins:2007ip,Nieves:2001fc}.

In our model, the light neutrino mass matrix depends on six real parameters: the magnitude and the phase of the modulus $\tau$, the overall neutrino scale $\kappa$, the parameter $\tilde\alpha_D$, and the magnitude and the phase of $\tilde\gamma_D$.
One can determine the viable parameter space of the model by fitting to the six neutrino oscillation parameters. To this end, we use the oscillation parameters determined from the global neutrino data fit~\cite{deSalas:2020pgw,Capozzi:2021fjo,Esteban:2024eli}. The best-fit values and the 3$\sigma$ ranges of the neutrino oscillation parameters are shown in Tab.~\ref{table:data_nufit}. 
\begin{table}[ht]
    \centering
    \begin{tabular}{|l|c|c|c|c|c|l|}
    \hline\hline
\multirow{2}{*}{Parameters}  & \multicolumn{2}{c|}{Normal ordering}  & \multicolumn{2}{c|}{Inverted ordering} \\\cline{2-5}
    &Best-fit value & $3\sigma$ range &  Best-fit value & $3\sigma$ range \\ \hline
$\sin^2\theta_{12}$ & $0.308_{-0.011}^{+0.012}$  & $0.275 - 0.345$ & $0.308_{-0.011}^{+0.012}$ & $0.275 - 0.345$ \\
$\sin^2\theta_{23}$ & $0.470_{-0.013}^{+0.017}$ & $0.435 - 0.585$ & $0.550_{-0.015}^{+0.012}$ & $0.440 - 0.584$\\
$\sin^2\theta_{13}$ & $0.02215_{-0.00058}^{+0.00056}$ & $0.02030 - 0.02388$ & $0.02231_{-0.00056}^{+0.00056}$ & $0.02060 - 0.02409$ \\
\hline 
$\Delta m_{\mathrm{sol}}^2/10^{-5}$ eV$^2$ & $7.49_{-0.19}^{+0.19}$ & $6.92- 8.05$ & $7.49_{-0.19}^{+0.19}$ & $6.92- 8.05$\\
        $\Delta m_{\mathrm{atm}}^2/10^{-3}$ eV$^2$ & $2.513_{-0.019}^{+0.021}$ & $2.451- 2.578$ & $2.484_{-0.020}^{+0.020}$ & $ 2.421-2.547$ \\
        \hline
        $\delta_{\mathrm{CP}}/^{\circ}$ & $212_{-41}^{+26}$ & $ 124- 364$  & $ 274_{-25}^{+22}$  & $201 - 335$ \\
\hline
\hline
\end{tabular}
    \caption{Best-fit values and their $3\sigma$ ranges for neutrino oscillation parameters obtained from NuFIT 6.0~\cite{Esteban:2024eli}.}
    \label{table:data_nufit}
\end{table}

In addition to the six neutrino oscillation parameters, the overall neutrino mass scale can be probed by various experiments. The sum of light neutrino masses, $\sum_i m_i$, is constrained by the cosmic microwave background measurements and the baryon acoustic oscillation data. From the analyses by the authors of~\cite{Vagnozzi:2017ovm,Planck:2018vyg,DiValentino:2019dzu,Jiang:2024viw}, one gets  $\sum_im_i\le0.12-0.52$ eV at 95\% confidence level (CL). The range in the upper limit reflects the uncertainty due to the assumed cosmology model. The effective neutrino mass, $m_{\nu_e}^{\rm eff} = \sqrt{\sum_i|U_{ei}^2|m_i}$, can be deduced from the end point of the beta decay spectrum. With the latest data on tritium beta decay, KATRIN has placed a constraint $m_{\nu_e}^{\rm eff}\le0.45$ ev at 90\% CL~\cite{Katrin:2024tvg}. Finally, the effective mass for neutrino-less double beta decay ($0\nu\beta\beta$), $m_{ee}$, is given by the 1-1 component of the light neutrino mass matrix in the basis where charged lepton mass is diagonal. The most stringent constraint on $m_{ee}$ is provided by the KamLAND-Zen experiment, with $m_{ee}\le0.036-0.156$ eV at 90\% CL~\cite{KamLAND-Zen:2022tow}. Here, the range in the upper bound reflects the uncertainty associated with the nuclear matrix element computation.

\section{Lepton Flavor Violation}
\label{sec:lfv}
The leptonic super potential in Eq.~\eqref{eq:superpotential} leads to lepton flavor violating interactions
\begin{equation}
    \mathcal{L}_{\rm LFV} =\frac{\sqrt{2}(U_L^\dagger M_D^\dagger)_{iI}}{v}\frac{1}{\tan\beta}\left[ \bar \ell_i P_RN_IH^- + \frac{1}{\cos\beta}\bar{\ell}_iP_R\tilde{N}_I\tilde{H}_u^-\right] +\text{h.c.},
    \label{eq:lfvinteraction}
\end{equation}
where $\ell_i$ is the charged lepton in the mass eigenbasis, $\tan\beta=v_u/v_d$, and $H^- = \cos\beta H_u^- - \sin\beta H_d^-$ is the physical charged Higgs. The first term in Eq.~\eqref{eq:lfvinteraction} is the LFV interaction involving the charged Higgs and the heavy neutrino, while the second term involves the super partners. Note that in the large $\tan\beta$ limit, the super partners mediated LFV dominate over those of regular particles. In principle, $\tilde{N}_{1,2}$ can mix. However, for simplicity we will assume they do not mix. Similarly, we will also assume that $\tilde{H}_u^+$ does not mix with the other charginos.  

In general, one would also expect LFV from the soft SUSY breaking terms. Even if one forbid LFV at the SUSY breaking scale, the renormalization group (RG) running will induce LFV interactions at low scale proportional to $M_D^\dagger M_D^{}/(8\pi^2v_u^2)$~\cite{Casas:2001sr}. Such an effect is sub-leading compare to the LFV interactions in Eq.~\eqref{eq:lfvinteraction}. Thus, for simplicity, we will neglect the RG induced LFV in our analysis.

The LFV interactions in Eq.~\eqref{eq:lfvinteraction} lead to radiative decays $\ell_i\to\ell_j\gamma$. Such a process is generated at one-loop level via penguin diagrams with neutral and charged scalars in the loop. In our scenario, we have~\cite{Hisano:1995cp,Arganda:2005ji}
\begin{equation}
    \Gamma(\ell_i\to\ell_j\gamma) = \frac{\alpha m_{\ell_i}^5}{1024\pi^4 }\left|\frac{(U_L^\dagger M_D^\dagger)_{jI}(M_D^{}U_L^{})_{Ii}}{v^2\tan^2\beta}\left[\frac{1}{m_{H^+}^2}f\left(\frac{m_{N_I}^2}{m_{H^+}^2}\right) - \frac{\sec^2\beta}{m_{\tilde{N}_I^2}}g\left(\frac{m_{\tilde{H}_u^+}^2}{m_{\tilde{N}_I}^2}\right)\right]\right|^2,
    \label{eq:lfvpartialwidth}
\end{equation}
where we have neglected terms suppressed by $m_{\ell_j}/m_{\ell_i}$. In the above equation, the sum on $I=1,2$ is implicit and the loop functions $f$ and $g$ are given by 
\begin{align}
    f(x) &= \frac{1-6x+3x^2+2x^3-6x^2\ln x}{6(1-x)^4},\\
    g(x) &= \frac{2+3x-6x^2+x^3+6x\ln x}{6(1-x)^4}.
\end{align}

The decay $\mu\to e\gamma$ provides the tightest constrain on LFV interaction. Such a process has been search for at the MEG and MEG II experiments. The combination of MEG and MEG II results give $Br(\mu\to e\gamma)\le3.1\times10^{-13}$ at 90\% CL~\cite{MEG:2016leq,MEGII:2023ltw}. The decays $\tau\to e\gamma,\mu\gamma$ are less constraining. These processes have been searched for by BaBar and Belle. The strongest upper bounds on $\tau\to e\gamma$ and $\tau\to\mu\gamma$ are obtained by BaBar and Belle respectively, with $Br(\tau\to e\gamma)\le 3.3\times10^{-8}$~\cite{BaBar:2009hkt} and $Br(\tau\to\mu\gamma)\le4.2\times10^{-8}$ at 90\% CL~\cite{Belle:2021ysv}.

\section{Numerical result}
\label{sec:results}
In this section, we identify the region of our model parameter space consistent with neutrino oscillation data at the 3$\sigma$ level, see Tab.~\ref{table:data_nufit}. Recall that in our set up, we have 2 complex parameters: the modulus $\tau$ and the ratio of Yukawa couplings $\tilde\gamma_D$. In addition, we have two real parameters: the overall neutrino mass scale $\kappa$, and the Yukawa couplings ratio $\tilde\alpha_D$. In our scan, $\tau$ is varied within the fundamental domain defined by 
\begin{equation}
    \text{Re}(\tau)\le1/2,\quad\text{and}\quad|\tau|\ge 1.
\end{equation} 
The range of the neutrino mass scale is taken to be
\begin{equation}
    0.1\text{ meV} \le \kappa \le 10 \text{ eV},
\end{equation}
while the ratio of the Yukawa couplings are taken to be in the range
\begin{equation}
    \tilde\alpha_D,|\tilde\gamma_D| \in \left[10^{-3}, 10^3\right],
\end{equation}
to mimic the hierarchy in the Yukawa couplings in the SM. Finally, the phase of $\gamma_D$ is varied within the range
\begin{equation}
    \phi_{\gamma_D} \in [0,2\pi].
\end{equation} 

\begin{figure}
    \centering
    \includegraphics[width=0.6\linewidth]{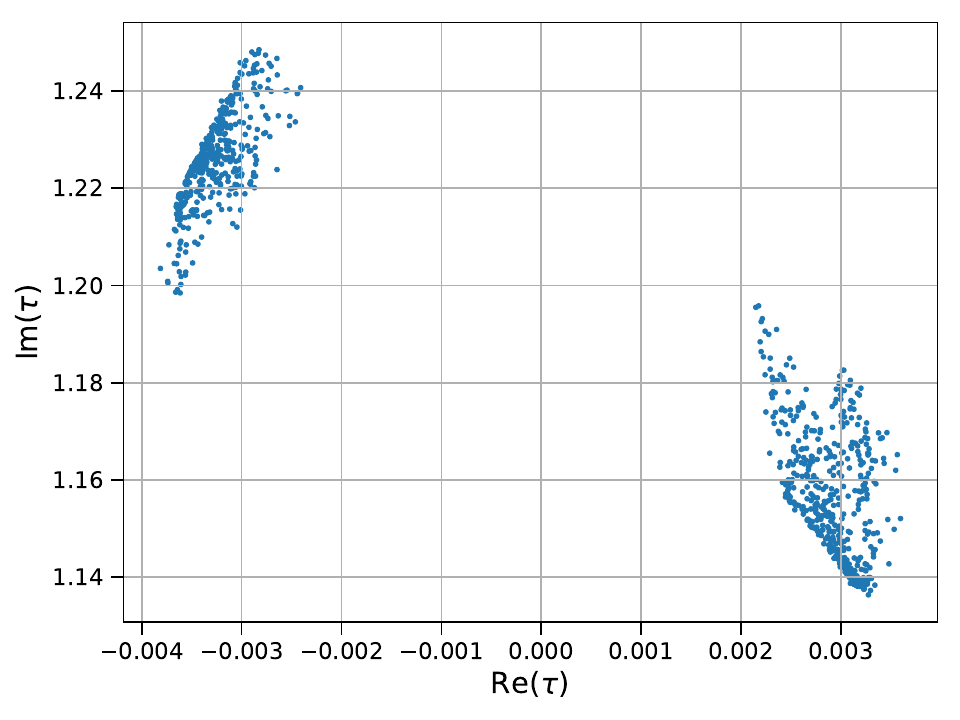}
    \caption{The space of modulus $\tau$ compatible with neutrino oscillation data. Notice the viable values of $\tau$ are clustered around $\tau\simeq1.2i$.}
    \label{fig:modulus}
\end{figure}

\begin{figure}
    \centering
    \includegraphics[width=0.6\linewidth]{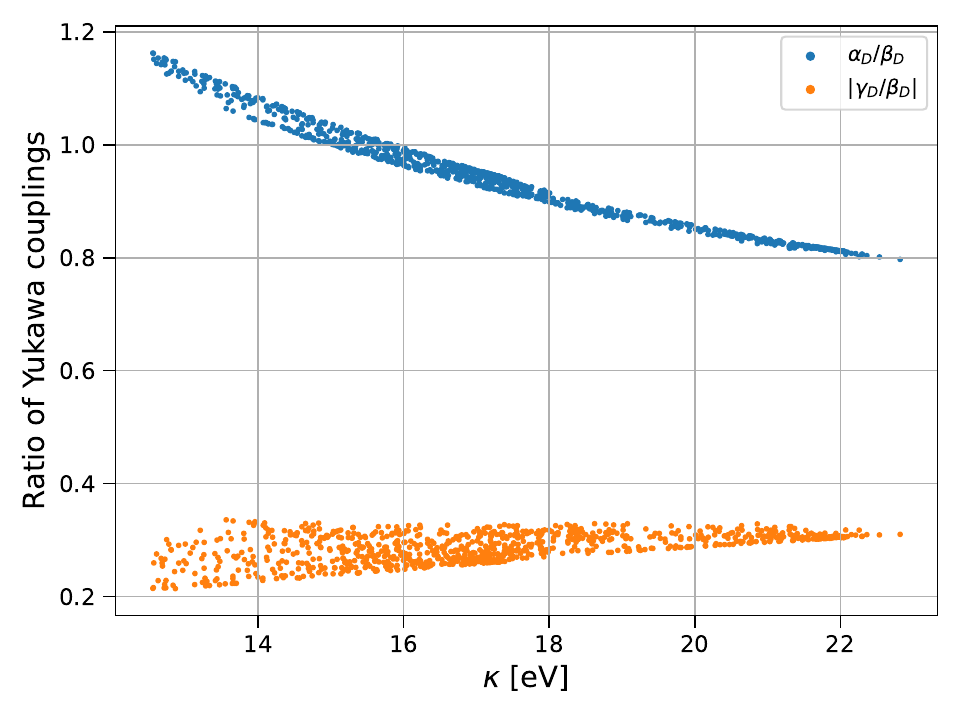}
    \caption{The relations between the overall neutrino mass scale $\kappa$ and the Yukawa coupling ratios compatible with neutrino oscillation data. Notice that $\alpha_D\simeq\beta_D$ while $|\gamma_D|$ is smaller by roughly an order of magnitude.}
    \label{fig:yukawa}
\end{figure}

From the scan, we find that our model can only accommodate IO neutrino masses. The viable parameter space is clustered around $\tau \simeq 1.2i$, see Fig.~\ref{fig:modulus}. The Yukawa coupling ratios $\alpha_D/\beta_D$ and $|\gamma_D/\beta_D|$ are strongly correlated with the overall neutrino scale $\kappa$, see Fig.~\ref{fig:yukawa}. Moreover, we find that $\alpha_D\simeq\beta_D$ while $|\gamma_D|$ is typically suppressed by an order of magnitude compared to $\alpha_D$ and $\beta_D$. Since the lightest neutrino mass in our model is vanishing, we have $m_3=0$. Thus, we get 
\begin{align}
    \sum_i m_i &\simeq 2\sqrt{\Delta m^2_{\rm atm}}\simeq100 \text{ meV},\\
    m_{\nu_e}^{\rm eff} &\simeq \phantom{2}\sqrt{\Delta m^2_{\rm atm}}\simeq \phantom{1}50 \text{ meV},
\end{align}
where we have dropped terms suppressed by $\sin^2\theta_{13}$ and $\Delta m^2_{\rm sol}/\Delta m^2_{\rm atm}$. The $\sum_i m_i$ predicted by our model lies just below the tightest constraint from cosmology $\sum_i m_i\le 130$ meV, while the effective mass $m_{\nu_e}^{eff}$ is an order of magnitude below the KATRIN bound, $m_{\nu_e}^{eff}\le450$ meV. The effective mass $m_{ee}$ in our scenario, however, lies in the range 38 meV $\le m_{ee}\le48$ meV. It lies just above the tightest KamLAND-ZEN constraint, $m_{ee}\le36-156$ meV. However, one cannot draw a definite conclusion from our $m_{ee}$ result due to uncertainty in nuclear matrix element involved in deducing the KamLAND-Zen bound.

\begin{figure}
    \centering
    \includegraphics[width=0.7\linewidth]{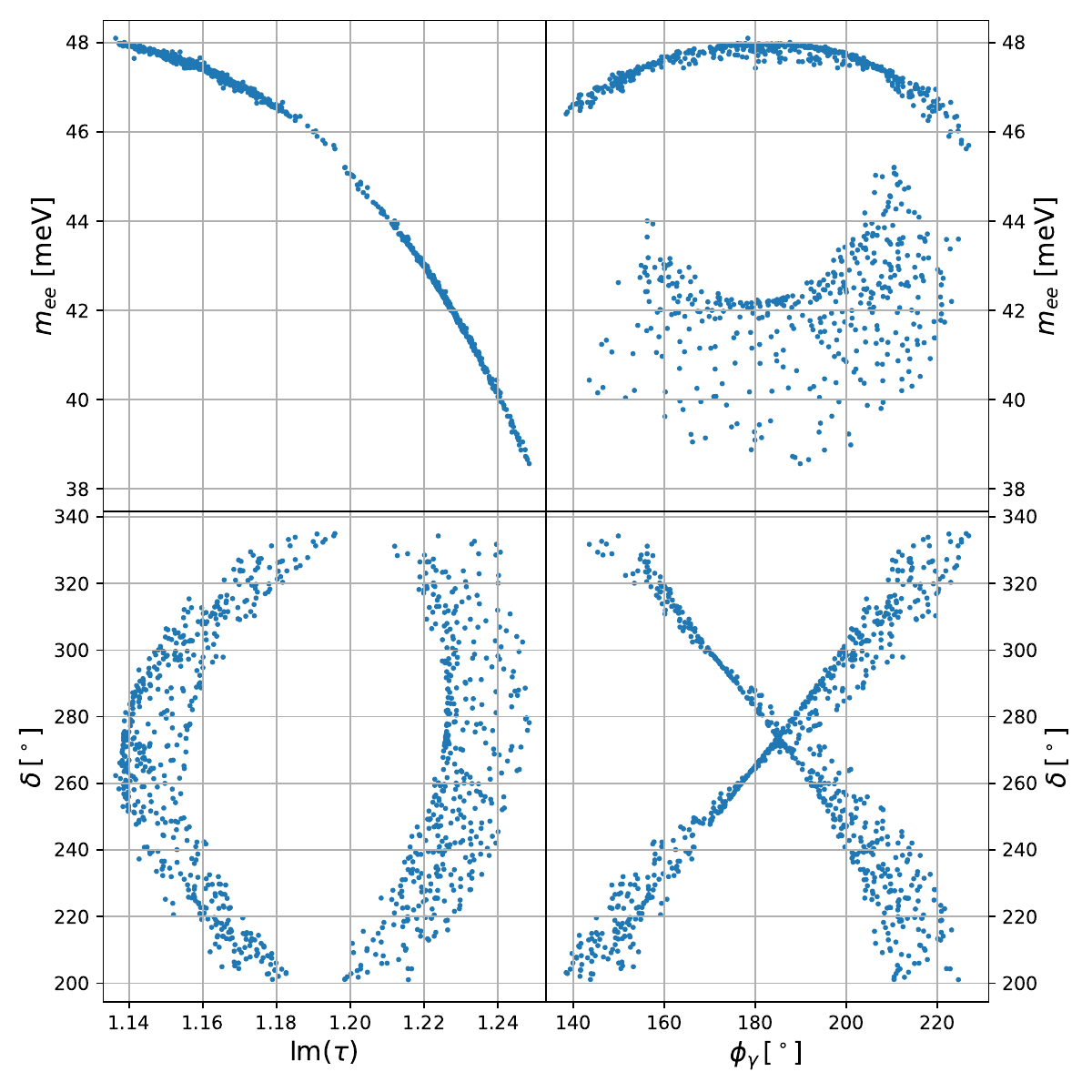}
    \caption{The relations between the free parameters Im$(\tau)$ and $\phi_{\gamma_D}$, and their corresponding $CP$ violating observables $\delta$ and $m_{ee}$. There is a strong correlation between Im$(\tau)$ and $m_{ee}$.}
    \label{fig:cpv}
\end{figure}

In our set up, $CP$ violation arises from the imaginary part of the modulus $\tau$ and the phase of the Yukawa coupling $\phi_{\gamma_D}$. From our scan, the phase $\phi_{\gamma_D}$ is in the range $2.4\le\phi_{\gamma_D}\le3.9$. These two $CP$ violating parameters are strongly correlated with the Dirac phase $\delta$ characterizing neutrino oscillation, and the effective mass $m_{ee}$, see Fig.~\ref{fig:cpv}. It should be noted that, in our scenario, $201^\circ\lesssim\delta\lesssim335^\circ$, which covers the full range of $\delta$ allowed by neutrino oscillation data, see Tab.~\ref{table:data_nufit}. 

\begin{figure}
    \centering
    \includegraphics[width=0.6\linewidth]{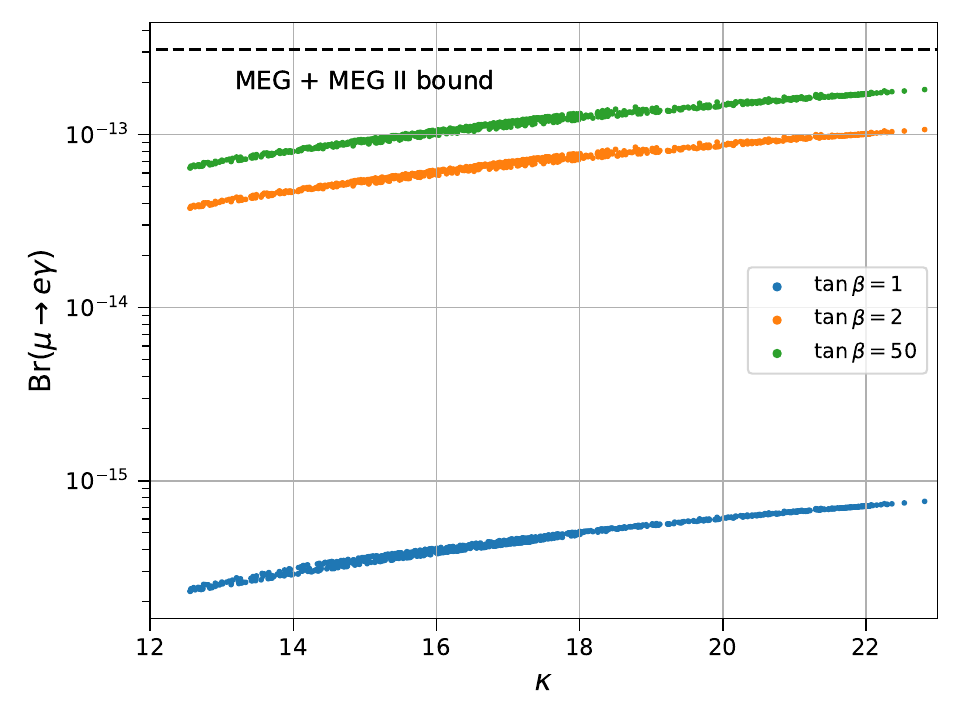}
    \caption{The Br($\mu\to e\gamma$) as a function of $\kappa$ for the benchmark scenario $M=M_{H^+}=M_{\tilde{N}_I}=M_{\tilde H_u^+}=1$ TeV and $\mu_S=10$ eV. The blue, orange, and green points correspond to $\tan\beta = 1$, 2 and 50 respectively.}
    \label{fig:meg}
\end{figure}

The Yukawa interaction responsible for generating neutrino masses also induces radiative LFV decays $\mu\to e\gamma$, $\tau\to e\gamma$ and $\tau\to\mu\gamma$ at the one-loop level. The partial decay widths for these modes are given by Eq.~\eqref{eq:lfvpartialwidth}. In addition to the Yukawa couplings $\alpha_D$, $\beta_D$, $\gamma_D$, the partial widths also depend on $\tan\beta$ and the masses of heavy particles running in the loop. The masses of the heavy neutrinos $N^c$ depend on the parameters $M$ and the modular forms $Y_\pm$. However, the masses of other heavy particles are arbitrary. In our analysis, the ratios $\alpha_D/\beta_D$ and $\gamma_D/\beta_D$ are determined from the fit to the neutrino oscillation data. The coupling $\beta_D$, however, cannot be directly determined from the fit. Instead, $\beta_D$ is related to $\tan\beta$, $M$, and $\mu_S$ via the overall neutrino scale $\kappa$, see Eq.~\eqref{eq:kappa}. For the given values of $\tan\beta$ and $M$, we get $\tan\beta^2\sim1/\mu_S$. Thus, from Eq.~\eqref{eq:lfvpartialwidth} one sees that the radiative LFV partial decay width scales with $\mu_S$ as 
\begin{equation}
    \Gamma[\mu_S'] = \frac{\mu_S^2}{\mu_S^{\prime2}}\Gamma[\mu_S].
\end{equation}

As an illustrative example, consider the benchmark scenario where $M=m_{H^+}=m_{\tilde N_i} = m_{\tilde{H}_u^+}=1$ TeV and $\mu_S=10$ eV. For $\tan\beta\gtrsim2$, the $\ell\to\ell'\gamma$ partial decay width is dominated by the neutral scalar contribution, i.e. the second term in Eq.~\eqref{eq:lfvpartialwidth}. In this case, our model predicts Br$(\mu\to e\gamma)$ in the order $10^{-14}-10^{-13}$, just below the current combined MEG and MEG II bound, see Fig.~\ref{fig:meg}. For $\tan\beta=1$, due to cancellation between the neutral and charged scalar loop, see Eq.~\eqref{eq:lfvpartialwidth}, the Br$(\mu\to e\gamma)$ is further suppressed by approximately two orders of magnitude. 
For the radiative LFV decays of the $\tau$, our benchmark scenario predicts their branching fractions to be of the same order as Br($\mu\to e\gamma$). Thus, they are well below the current experimental limits.

\section{Conclusion and discussion}
\label{sec:con}
In this work, we construct the minimal inverse seesaw neutrino mass model compatible with modular $S_3$ symmetry.  In our model, the MSSM is extended by a pair of electroweak singlet $S_3$ doublet leptons $N^c$ and $S$. In our framework, the leptonic Yukawa couplings are modular forms, thereby endowing them with transformation properties under the modular $S_3$ group and enhancing predictability. Due to the minimal structure of the model, our light neutrino mass matrix is rank two. Hence, the lightest neutrino is massless. As a result, the sum of light neutrino masses, $\sum_i m_i$, and the effective mass for the endpoint of beta decay spectrum, $m_{\nu_e}^{\rm eff}$, are completely determined in terms of neutrino oscillation parameters.  

From our analysis, we find our model only accommodates IO neutrino masses. As a result, $\sum_i m_i\simeq 100$ meV, just below the Planck constraint $\sum_i m_i\le130$ meV at 95\% CL. However, it is within reach of the next generation cosmological measurement such as the Simon Observatory whose projected sensitivity is $\sum_i m_i\le40$ meV at 95\% CL~\cite{SimonsObservatory:2019qwx}. Moreover, the effective mass is $m_{\nu_e}^{\rm eff}\simeq 50$ meV, which is about an order of magnitude below the current KATRIN bound. It is also well below the expected reach, $m_{\nu_e}^{\rm eff}\simeq 200$ at 90\% CL, of both the Holmes and the KATRIN experiments~\cite{Alpert:2014lfa,KATRIN:2019yun}. The effective mass $m_{ee}$, unlike $\sum_i m_i$ and $m_{\nu_e}^{\rm eff}$, is not fixed by the neutrino oscillation parameters. In fact, it is strongly correlated with the free Im$(\tau)$. Our parameter scan finds $m_{ee}=38-48$ meV. This range is probed by the KamLAND-Zen experiment, whose sensitivity is $m_{ee}\le36-156$ meV. However, due to uncertainty in evaluating nuclear matrix elements, one should not yet draw any definite conclusion on the validity of the model based on the KamLAND-Zen result. The next generation $0\nu\beta\beta$ experiment, such as the phase-II of the LEGEND experiment whose project sensitivity is $m_{ee}\le 13-29$ meV at 90\% CL~\cite{LEGEND:2017cdu}, could decisively discover or rule out our scenario. 

The Yukawa couplings that generate neutrino masses through the inverse seesaw mechanism also lead to radiative LFV decays $\ell\to\ell'\gamma$. The partial decay widths are related to the overall neutrino mass scale $\kappa$. For the benchmark scenario with $M=m_{H^+}=m_{\tilde N_i} = m_{\tilde{H}_u^+}=1$ TeV and $\mu_S=10$ eV, the branching ratio $\mu\to e\gamma$ is below the constraint from MEG and MEG II for  $\tan\beta=1-50$. The decay mode $\tau\to e\gamma$ and $\tau\to\mu\gamma$ are well below their respective constraints as well. Note that the higher the $\mu_S$, the smaller the branching ratios.  Hence, in our model, there is always a part of parameter space which is compatible with LFV constraints.  

\appendix

\section{The $S_3$ modular group}
\label{sec:S3_symmetry}
The group $S_3$ admits three irreducible representations: the singlet $\textbf{1}$, the pseudo-singlet $\textbf{1}'$ and the doublet $\textbf{2}$. In our model, however, we make use only of the singlet and doublet representations. In the case of $S_3$, there are two linearly independent lowest weight ($k=2$) modular forms. They transform in the doublet representation
\begin{equation}
    Y_\textbf{2}^{(2)}=\begin{pmatrix}Y_1(\tau)\\Y_2(\tau)\end{pmatrix},
\end{equation}
where $\tau$ is the complex modulus, the subscript and the superscript denote the $S_3$ representation and the modular weight respectively. The modular forms $Y_1$ and $Y_2$ are given in term of the Dedekind eta function ($\eta(\tau)$)~\cite{Kobayashi:2018vbk}. However, in practice, they can be expanded as
\begin{align} 
Y_1(\tau) &= \frac 18 + 3q^2 + 3q^4 + 12 q^6 + 3q^8 \cdots , \\
Y_2(\tau) &= \sqrt 3 q (1+ 4 q^2 + 6 q^{4} + 8 q^{6} \cdots  ),  \label{eq:modular func_S3_expand}
\end{align}
where $q \equiv e^{i\pi\tau}$.

For model building, one also need the tensor products of irreducible representations of $S_3$. In our case, only the tensor product of two doublets is needed. It is given by~\cite{Ishimori:2010au}
\begin{equation}
    \begin{pmatrix}x_1\\x_2\end{pmatrix}_{\textbf{2}}\otimes\begin{pmatrix}x_1\\x_2\end{pmatrix}_{\textbf{2}} = (x_1y_1 + x_2y_2)_{\textbf{1}} \oplus (x_1y_2-x_2y_1)_{\textbf{1}'} \oplus \begin{pmatrix}x_2y_2-x_1y_1\\x_1y_2+x_2y_1\end{pmatrix}_{\textbf{2}}.
\end{equation}
The tensor product allow one to obtain the higher weight modular forms $Y_{\textbf{a}}^{(k)}$ where $\textbf{a}$ is the $S_3$ representation and $(k)$ is the modular weight. Specifically, we have
\begin{equation}
    Y_{\textbf{1}}^{(4)} = Y^2_1+Y^2_2,\quad
    Y_{\textbf{2}}^{(4)} = \begin{pmatrix}Y^2_2-Y_1^2\\2Y_1Y_2\end{pmatrix}.
\end{equation}


\begin{acknowledgments}
MKB and PI acknowledge support from the NSRF via the Program Management Unit for Human Resources \& Institutional Development, Research, and Innovation [grant no. B13F670070]. The work of PU is supported in part by the National Science, Research and Innovation Fund (NSRF) for the fiscal year 2025 via Srinakharinwirot University  under Grant No. 046/2568. MKB and PU also acknowledge the National Science and Technology Development Agency, National e-Science Infrastructure Consortium, Chulalongkorn University, and the Chulalongkorn Academic Advancement into Its 2nd Century Project (Thailand) for providing computing infrastructure that has contributed to the results reported within this paper. CP is supported by Fundamental Fund 2568 of Khon Kaen University and Research Grant for New Scholar, Office of the Permanent Secretary, Ministry of Higher Education, Science, Research and Innovation under contract no. RGNS 64-043.
\end{acknowledgments}


\bibliographystyle{utcaps_mod}
\bibliography{s3_inverse}

\end{document}